\documentclass[12pt]{article}
\usepackage{fullpage}
\usepackage{hyperref}

\usepackage{graphicx}
\usepackage{pythontex}
\newcommand{\erdosrenyi}{Erd\H{o}s--R\'{e}nyi}

\let\temptt\texttt
\renewcommand\texttt[1]{\mbox{\temptt{#1}}}

\newcommand{\footremember}[2]{%
    \footnote{#2}
    \newcounter{#1}
    \setcounter{#1}{\value{footnote}}%
}
\newcommand{\footrecall}[1]{%
    \footnotemark[\value{#1}]%
} 

\title{EoN (Epidemics on Networks): a fast, flexible Python package for simulation, analytic approximation, and analysis of epidemics on networks}

\author{Joel C. Miller\footnote{La Trobe University} \footremember{IDM}{Institute for Disease Modeling}%
\and Tony Ting\footrecall{IDM}}
\date{}

\begin{document}
\maketitle
\begin{abstract}
We provide a description of the Epidemics on Networks (EoN) python package designed for studying disease spread in static networks.   The package consists of over $100$ methods available for users to perform stochastic simulation of a range of different processes including SIS and SIR disease, and generic simple or comlex contagions.

\begin{center}
This paper is available in a shorter form without examples at\\ \href{https://joss.theoj.org/papers/10.21105/joss.01731.pdf}{https://joss.theoj.org/papers/10.21105/joss.01731.pdf}
\end{center}
\end{abstract}

\section*{Introduction}

EoN (EpidemicsOnNetworks) is a pure-python package designed to assist studies of
infectious processes spreading through networks.  It originally rose out of the 
book \emph{Mathematics of Epidemics on Networks}~\cite{kiss:EoN}, and now consists of over 
100 user-functions.

EoN provides a set of tools for

\begin{itemize}
\item Susceptible-Infected-Susceptible (SIS) and Susceptible-Infected-Recovered 
(SIR) disease
\begin{itemize}
  \item Stochastic simulation of disease spread in networkx graphs
  \begin{itemize}
    \item continuous time Markovian
    \item continuous time nonMarkovian
    \item discrete time
    \end{itemize}
  \item Numerical solution of over 20 differential equation models, including
\begin{itemize}
    \item individual and pair-based models
    \item pairwise models
    \item edge-based compartmental models
    \end{itemize}
  \end{itemize}
\item Stochastic simulation of a wide range of Simple and Complex contagions
\item Visualization and analysis of stochastic simulations
\end{itemize}

These algorithms are built on the networkx package~\cite{hagberg2008exploring}.
EoN's documentation is maintained at 
\begin{center}
\href{https://epidemicsonnetworks.readthedocs.io/en/latest/}{https://epidemicsonnetworks.readthedocs.io/en/latest/}
\end{center}
including numerous examples at 
\begin{center}
\href{https://epidemicsonnetworks.readthedocs.io/en/latest/Examples.html}{https://epidemicsonnetworks.readthedocs.io/en/latest/Examples.html}.
\end{center}
In this 
paper we provide brief descriptions with examples of a few of EoN's tools.
The examples shown are intended to demonstrate the ability of the tools.  The 
online documentation gives more detail about how to use them.

We model spreading processes on a contact network.  In this context, many
mathematicians and physicists are accustomed to thinking of individuals as 
nodes with their potentially infectious partnerships as edges.  However, for 
those who come from other backgrounds this abstraction may be less familiar.  

Therefore, we will describe a contact network along which an infections 
process spreads as consisting of ``individuals'' and ``partnerships'' rather than 
``nodes'' and ``edges''.  This has an additional benefit because in the simple 
contagion algorithm (described later), we need to define some other networks 
whose nodes represent possible statuses and whose edges represent transitions 
that can occur.  Referring to ``individuals'' and ``partnerships'' when discussing
the process spreading on the contact network makes it easier to avoid 
confusion between the different networks.

We start this paper by describing the tools for studying SIS and SIR disease 
through stochastic simulation and differential equations models.  Then we 
discuss the simple and complex contagions, including examples showing how
the simple contagion can be used to capture a range of standard disease
models.  Finally we demonstrate the visualization tools.

\section*{SIR and SIS disease}

\subsection*{Stochastic simulation}
The stochastic SIR and SIS simulation tools allow the user to investigate
standard SIS and SIR dynamics (SEIR/SIRS and other processes are addressed 
within the simple contagion model):

\begin{itemize}
\item Markovian SIS and SIR simulations  (\texttt{fast\_SIS}, \texttt{Gillespie\_SIS}, \texttt{fast\_SIR}, and \texttt{Gillespie\_SIR}).
\item non-Markovian SIS and SIR simulations (\texttt{fast\_nonMarkovian\_SIS} and \texttt{fast\_nonMarkovian\_SIR}).
\item discrete time SIS and SIR simulations where infections last a single time step 
  (\texttt{basic\_discrete\_SIS}, \texttt{basic\_discrete\_SIR}, and \texttt{discrete\_SIR}).
\end{itemize}

For both Markovian and non-Markovian methods it is possible for the transition 
rates to depend on intrinsic properties of 
individuals and of partnerships.

The continuous-time stochastic simulations have two different implementations: a 
Gillespie implementation~\cite{gillespie1977exact,doob1945markoff} and an Event-driven
implementation.  Both approaches are efficient.  They have similar speed if the 
dynamics are Markovian (depending on the network and disease parameters either
may be faster than the other), but the event-driven implementation can also handle 
non-Markovian dynamics.  In earlier versions, the event-driven simulations were 
consistently faster than the Gillespie simulations, and thus they are named 
\texttt{fast\_SIR} and \texttt{fast\_SIS}.  The Gillespie simulations have since reached
comparable speed using ideas from~\cite{holme2014model} and~\cite{cota2017optimized}.

The algorithms can typically handle an SIR epidemic spreading on 
hundreds of thousands of individuals in well under a minute on a laptop.  The 
SIS versions are slower because the number of events that happen is often much
larger in an SIS simulation.

\subsubsection*{Examples}

To demonstrate these, we begin with SIR simulations on an \erdosrenyi{} network
having a million individuals (in an \erdosrenyi{} network each individual has 
identical probability of independently partnering with any other individual in 
the population).

\begin{pyverbatim}
import networkx as nx
import EoN
import matplotlib.pyplot as plt


N = 10**6  #number of individuals
kave = 5    #expected number of partners
print('generating graph G with {} nodes'.format(N))
G = nx.fast_gnp_random_graph(N, kave/(N-1)) #Erdo''s-Re'nyi graph
    
rho = 0.005 #initial fraction infected
tau = 0.3   #transmission rate
gamma = 1.0 #recovery rate

print('doing event-based simulation')
t1, S1, I1, R1 = EoN.fast_SIR(G, tau, gamma, rho=rho)
#instead of rho, we could specify a list of nodes as initial_infecteds, or
#specify neither and a single random node would be chosen as the index case.

print('doing Gillespie simulation')
t2, S2, I2, R2 = EoN.Gillespie_SIR(G, tau, gamma, rho=rho)

print('done with simulations, now plotting')
plt.plot(t1, I1, label = 'fast_SIR')
plt.plot(t2, I2, label = 'Gillespie_SIR')
plt.xlabel('$t$')
plt.ylabel('Number infected')
plt.legend()
plt.show()
\end{pyverbatim}

This produces a (stochastic) figure like

\includegraphics[width=0.8\textwidth]{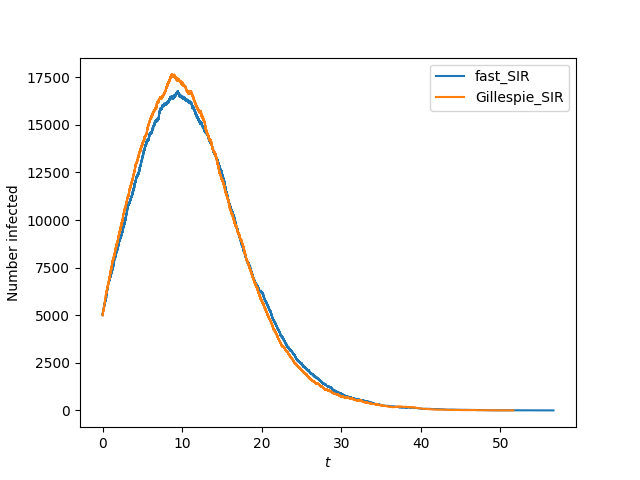}    
    
The run-times of \texttt{fast\_SIR} and \texttt{Gillespie\_SIR} are both comparable to the
time taken to generate the million individual network \texttt{G}.  The epidemics 
affect around 28 percent of the population.  The differences between the simulations 
are entirely due to stochasticity.

We can perform similar simulations with an SIS epidemic.  Because SIS epidemics
take longer to simulate, we use a smaller network and specify the optional 
\texttt{tmax} argument defining the maximum stop time (by default \texttt{tmax=100}).

\begin{pyverbatim}
import networkx as nx
import EoN
import matplotlib.pyplot as plt


N = 10**5   #number of individuals
kave = 5    #expected number of partners
print('generating graph G with {} nodes'.format(N))
G = nx.fast_gnp_random_graph(N, kave/(N-1)) #Erdo''s-Re'nyi graph
    
rho = 0.005 #initial fraction infected
tau = 0.3   #transmission rate
gamma = 1.0 #recovery rate
print('doing Event-driven simulation')
t1, S1, I1 = EoN.fast_SIS(G, tau, gamma, rho=rho, tmax = 30)
print('doing Gillespie simulation')
t2, S2, I2 = EoN.Gillespie_SIS(G, tau, gamma, rho=rho, tmax = 30)

print('done with simulations, now plotting')
plt.plot(t1, I1, label = 'fast_SIS')
plt.plot(t2, I2, label = 'Gillespie_SIS')
plt.xlabel('$t$')
plt.ylabel('Number infected')
plt.legend()
plt.show()
\end{pyverbatim}

This produces a (stochastic) figure like

\includegraphics[width=0.8\columnwidth]{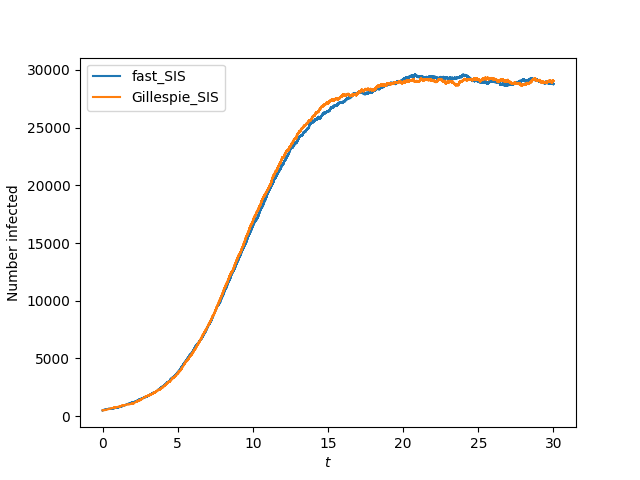}

We now consider an SIR disease spreading with non-Markovian dynamics.  We assume
that the infection duration is gamma distributed, but the transmission rate is
constant (yielding an exponential distribution of time to transmission).

This follows~\cite{vizi2019monotonic}.

\begin{pyverbatim}
import networkx as nx
import EoN
import matplotlib.pyplot as plt
import numpy as np

def rec_time_fxn_gamma(u):
    return np.random.gamma(3,0.5) #gamma distributed random number

def trans_time_fxn(u, v, tau):
    if tau >0:
        return np.random.exponential(1./tau)
    else:
        return float('Inf')
        
N = 10**6  #number of individuals
kave = 5    #expected number of partners
print('generating graph G with {} nodes'.format(N))
G = nx.fast_gnp_random_graph(N, kave/(N-1)) #Erdo''s-Re'nyi graph
tau = 0.3

for cntr in range(10):
    print(cntr)
    print('doing Event-driven simulation')
    t, S, I, R = EoN.fast_nonMarkov_SIR(G, trans_time_fxn = trans_time_fxn,
                                        rec_time_fxn = rec_time_fxn_gamma, 
                                        trans_time_args = (tau,))

    #To reduce file size and make plotting faster, we'll just plot 1000
    #data points.  It's not really needed here, but this demonstrates
    #one of the available tools in EoN.

    subsampled_ts = np.linspace(t[0], t[-1], 1000)
    subI, subR = EoN.subsample(subsampled_ts, t, I, R) 
    print('done with simulation, now plotting')
    plt.plot(subsampled_ts, subI+subR)

plt.xlabel('$t$')
plt.ylabel('Number infected or recovered')
plt.show()                                                            
\end{pyverbatim}    
This produces a (stochastic) figure like

\includegraphics[width=0.8\columnwidth]{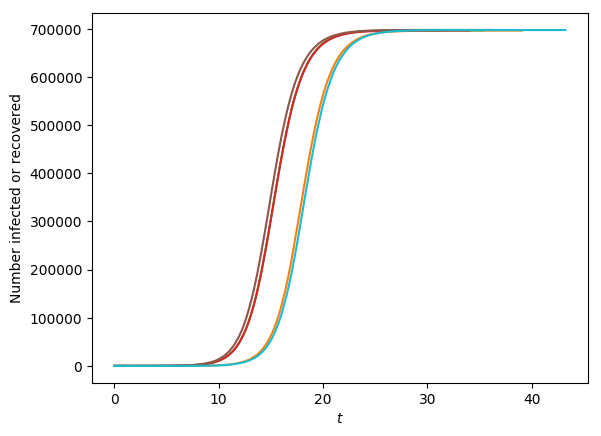}

\subsection*{Differential Equations Models}

EoN also provides a set of tools to numerically solve approximately 20 differential equations
models for SIS or SIR disease spread in networks.  The various models use different
information about the network to make deterministic predictions about
the number infected at different times.  These use the Scipy integration tools.
The derivations of the models and explanations of their simplifying assumptions
are described in~\cite{kiss:EoN}.

Depending on the model, we need different information about the network structure.
The algorithms allow us to provide the information as inputs.  However, there
is also a version of each model which takes a network as an input 
instead and then measures the network properties.

\subsubsection*{Examples}

We demonstrate an SIS pairwise model and an SIR edge-based compartmental model.

Our first example uses an SIS homogeneous pairwise model (section 4.3.3 of~\cite{kiss:EoN}).  This model uses the average degree of the population and then 
attempts to track the number of [SI] and [SS] pairs.   We assume a network 
with an average degree of 20.  The initial condition is that a fraction $\rho$ 
(\texttt{rho}) of the population is infected at random.  

\begin{pyverbatim}
import networkx as nx
import EoN
import matplotlib.pyplot as plt

N=10000
gamma = 1
rho = 0.05
kave = 20
tau = 2*gamma/ kave
S0 = (1-rho)*N
I0 = rho*N
SI0 = (1-rho)*kave*rho*N
SS0 = (1-rho)*kave*(1-rho)*N
t, S, I = EoN.SIS_homogeneous_pairwise(S0, I0, SI0, SS0, kave, tau, gamma,
                            tmax=10)
plt.plot(t, S, label = 'S')
plt.plot(t, I, label = 'I')
plt.xlabel('$t$')
plt.ylabel('Predicted numbers')
plt.legend()
plt.show()
\end{pyverbatim}
This produces 

\includegraphics[width=0.8\columnwidth]{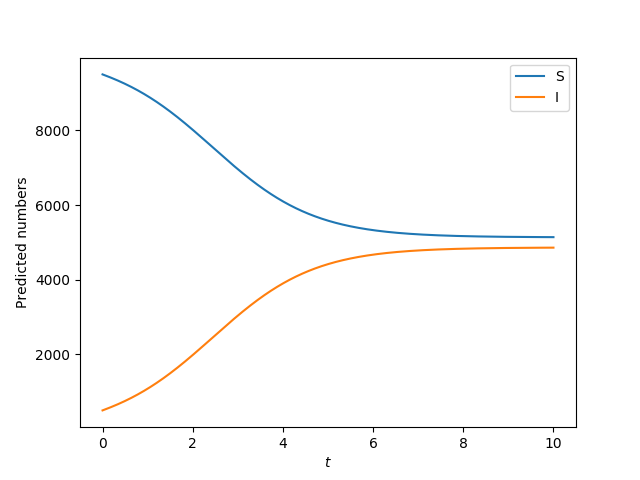}

For ease of comparison with simulation and consistency with existing literature,
the output of the model should be interpreted in terms of an expected number of individuals
in each status, which requires that our values scale with \texttt{N}.  So all of 
the initial conditions have a factor of \texttt{N}.  
If we are interested in proportion, we could arbitrarily set \texttt{N=1}, 
and then our solutions would give us the proportion of the population in each 
status.

Our second example uses an Edge-based compartmental model for an SIR disease 
(section 6.5.2 of~\cite{kiss:EoN} and also~\cite{miller:ebcm_overview,
miller:initial_conditions}).
This model incorporates information about the degree distribution (i.e., how the
number of partners is distributed), but assumes that the partnerships are selected
as randomly as possible given this distribution.  The model requires we define a
``generating function'' $\psi(x)$ which is equal to the sum 
$\sum_{k=0}^\infty S_k(0) x^k$ where
$S_k(0)$ is the proportion of all individuals in the population who both have $k$
partners and are susceptible at $t=0$.  It also requires the derivative
$\psi'(x)$ as well as $\phi_S(0)$, the probability
an edge from a susceptible node connects to another susceptible node at time 0.
By default, it assumes there are no recovered individuals at time $0$.

If the population has a Poisson degree distribution with mean \texttt{kave} and the 
infection is introduced by randomly infecting a proportion $\rho$ of the population
at time 0, then $\psi(x) = (1-\rho) e^{-\langle k\rangle (1-x)}$, 
$\ \psi'(x) = (1-\rho)\langle k\rangle e^{-\langle k \rangle(1-x)}$
and $\phi_S(0) = 1-\rho$ where $\langle k \rangle$ denotes \texttt{kave}.  So we have

\begin{pyverbatim}
import networkx as nx
import EoN
import matplotlib.pyplot as plt
import numpy as np

gamma = 1
tau = 1.5
kave = 3
rho = 0.01
phiS0 = 1-rho
def psi(x):
    return (1-rho)* np.exp(-kave*(1-x))
def psiPrime(x):
    return (1-rho)*kave*np.exp(-kave*(1-x))

N=1

t, S, I, R = EoN.EBCM(N, psi, psiPrime, tau, gamma, phiS0, tmax = 10)

plt.plot(t, S, label = 'S')
plt.plot(t, I, label = 'I')
plt.plot(t, R, label = 'R')
plt.xlabel('$t$')
plt.ylabel('Predicted proportions')
plt.legend()
plt.show()
\end{pyverbatim}

This produces 

\includegraphics[width=0.8\columnwidth]{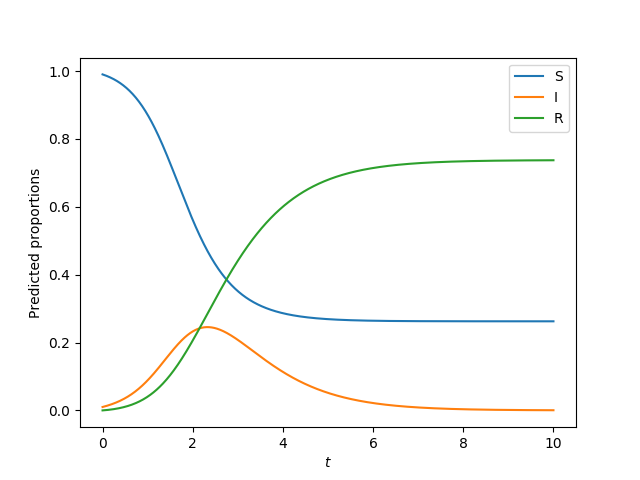}.

To be consistent with the other differential equations models, this EBCM implementation
returns the expected \emph{number} in each status, rather than the expected proportion.
Most of the literature on the EBCM approach~\cite{miller:ebcm_overview} focuses on expected proportion.
By setting \texttt{N=1}, we have found the proporitons of the population in each status.

\section*{Simple and Complex Contagions}

There are other contagious processes in networks which have received attention.
Many of these fall into one of two types, ``simple contagions'' and ``complex 
contagions''.

In a ``simple contagion'' an individual \texttt{u} may be induced to change status by
an interaction with its partner \texttt{v}.  This status change occurs with the same
rate regardless of the statuses of other partners of \texttt{u} (although the other
partners may cause \texttt{u} to change to another status first).  SIS and SIR 
diseases are special cases of simple contagions.

In a ``complex contagion'' however, we permit the rate at which \texttt{u} changes 
from one status to another to depend on the statuses of others in some more
complicated way.  Two infected individuals may cause a susceptible individual
to become infected at some higher rate than would result from them acting 
independently.  This is frequently thought to model social contagions where an
individual may only believe something if multiple partners believe it~\cite{centola:cascade}.  

The simple and complex contagions are currently implemented only in a 
Gillespie setting, and so they require Markovian assumptions.  Although they 
are reasonably fast, it would typically be feasible to make a bespoke algorithm
that runs significantly faster.

\section*{Simple contagions}

EoN provides a function \texttt{Gillespie\_simple\_contagion} which allows a user to 
specify the rules governing an arbitrary simple contagion.

Examples are provided in the online documentation, including

\begin{itemize}
\item SEIR disease (there is an exposed state before becoming infectious)
\item SIRS disease (recovered individuals eventually become susceptible again)
\item SIRV disease (individuals may get vaccinated) 
\item Competing SIR diseases (there is cross immunity)
\item Cooperative SIR diseases (infection with one disease helps spread the other)
\end{itemize}

The implementation requires the user to separate out two distinct ways that 
transitions occur: those that are intrinsic to an individual's current state
and those that are induced by a partner.  To help demonstrate, consider an 
``SEIR'' epidemic, where individuals begin susceptible, but when they interact 
with infectious partners they may enter an exposed state.  They remain in that 
exposed state for some period of time before transitioning into the infectious 
state independently of the status of any partner.
They remain infectious and eventually transition into the recovered state, again
independently of the status of any partner.  Here the ``E'' to ``I'' and ``I'' to ``R''
transitions are intrinsic to the individual's state, while the ``S'' to ``E'' 
transition is induced by a partner.  

To formalize this, we identify two broad types of transitions:

\begin{itemize}

\item \textbf{Spontaneous Transitions:}  Sometimes individuals change status without 
  influence from any other individual.  For example, an infected individual 
  may recover, or an exposed individual may move into the infectious class.  
  These transitions between statuses can be represented by a directed graph 
  \texttt{H} where the nodes are not the original individuals of the contact 
  network \texttt{G}, but rather the potential statuses individuals can take.  The 
  edges represent transitions that can occur, and we weight the edges by the 
  rate.  In the SEIR case we would need the graph \texttt{H} to have edges 
  \texttt{'E'}$\to$\texttt{'I'} and \texttt{'I'}$\to$\texttt{'R'}.  The edges would be weighted by the 
  transition rates.  Note \texttt{H} need not have a node \texttt{'S'} because 
  susceptible nodes do not change status on their own.

\item \textbf{Induced Transitions:} Sometimes individuals change status due to the 
  influence of a single partner.  For example in an SEIR model an infected 
  individual may transmit to a susceptible partner.  So an \texttt{('I', 'S')} pair
  may become \texttt{('I', 'E')}.  We can represent these transitions with a 
  directed graph \texttt{J}.  Here the nodes of \texttt{J} are pairs (tuples) of 
  statuses, representing potential statuses of individuals in a partnership.
  An edge represents a possible partner-induced transition.  In the SEIR case,
  there is only a single such transition, represented by the edge 
  \texttt{('I', 'S')} $\to$ \texttt{('I', 'E')} with a weight representing the transmission
  rate.  No other nodes are required in \texttt{J}.  An edge always represents the
  possibility that a node in the first state can cause the other node to change
  state.  So the first state in the pair remains the same.  The current 
  version does not allow for both nodes to simultaneously change states.
\end{itemize}  

\subsubsection*{Examples}

We first demonstrate a stochastic simulation of a simple contagion with an
SEIR example.  To demonstrate additional
flexibility we allow some individuals to have a higher rate of transitioning from 
\texttt{'E'} to \texttt{'I'} and some partnerships to have a higher transmission rate. 
This is done by adding weights to the contact network \texttt{G} which scale the 
rates for those individuals or partnerships.  The documentation discusses other ways we 
can allow for heterogeneity in transition rates.

Note that this process is guaranteed to terminate, so we can set \texttt{tmax} to 
be infinite.  Processes which may not terminate will require a finite value.
The default is 100.

\begin{pyverbatim}
import EoN
import networkx as nx
from collections import defaultdict
import matplotlib.pyplot as plt
import random

N = 100000
print('generating graph G with {} nodes'.format(N))
G = nx.fast_gnp_random_graph(N, 5./(N-1))

#We add random variation in the rate of leaving exposed class
#and in the partnership transmission rate.
#There is no variation in recovery rate.

node_attribute_dict = {node: 0.5+random.random() for node in G.nodes()}
edge_attribute_dict = {edge: 0.5+random.random() for edge in G.edges()}

nx.set_node_attributes(G, values=node_attribute_dict, 
                        name='expose2infect_weight')
nx.set_edge_attributes(G, values=edge_attribute_dict, 
                        name='transmission_weight')
#
#These individual and partnership attributes will be used to scale
#the transition rates.  When we define \texttt{H} and \texttt{J}, we provide the name
#of these attributes.

#More advanced techniques to scale the transmission rates are shown in
#the online documentation

H = nx.DiGraph() #For the spontaneous transitions
H.add_node('S') #This line is actually unnecessary.
H.add_edge('E', 'I', rate = 0.6, weight_label='expose2infect_weight')
H.add_edge('I', 'R', rate = 0.1)

J = nx.DiGraph() #for the induced transitions
J.add_edge(('I', 'S'), ('I', 'E'), rate = 0.1, 
            weight_label='transmission_weight')
IC = defaultdict(lambda: 'S')
for node in range(200):
    IC[node] = 'I'

return_statuses = ('S', 'E', 'I', 'R')

print('doing Gillespie simulation')
t, S, E, I, R = EoN.Gillespie_simple_contagion(G, H, J, IC, return_statuses,
                                        tmax = float('Inf'))

print('done with simulation, now plotting')
plt.plot(t, S, label = 'Susceptible')
plt.plot(t, E, label = 'Exposed')
plt.plot(t, I, label = 'Infected')
plt.plot(t, R, label = 'Recovered')
plt.xlabel('$t$')
plt.ylabel('Simulated numbers')
plt.legend()
plt.show()
\end{pyverbatim}

This produces a (stochastic) figure like 

\includegraphics[width=0.8\columnwidth]{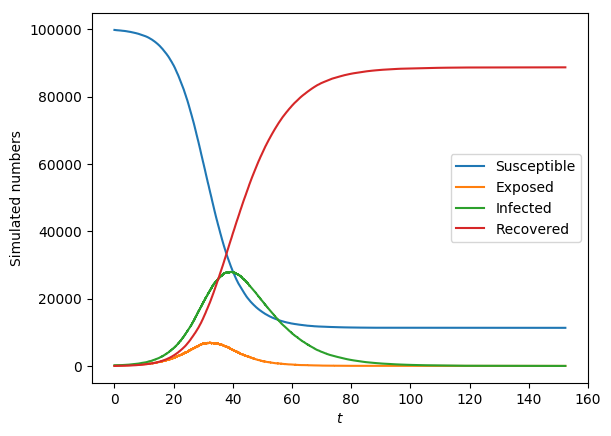}
    
Interaction between diseases can lead to interesting effects~\cite{grassberger2016phase,cui2017mutually,liu2018interactive,miller:compete}.
Now we show two cooperative SIR diseases.  In isolation, each of these diseases
would fail to start an epidemic.  However, together they can, and sometimes they
exhibit interesting osillatory behavior.  To help stimulate the oscillations, we
start with an asymmetric initial condition, though oscillations can be induced
purely by stochastic effects for smaller initial conditions.  To the best of
our knowledge, this oscillatory behavior has not been studied previously.

\begin{pyverbatim}
import EoN
import networkx as nx
from collections import defaultdict
import matplotlib.pyplot as plt

N = 300000
print('generating graph G with {} nodes'.format(N))
G = nx.fast_gnp_random_graph(N, 5./(N-1))

#In the below:
#'SS' means an individual susceptible to both diseases
#'SI' means susceptible to disease 1 and infected with disease 2
#'RS' means recovered from disease 1 and susceptible to disease 2.
#etc.

H = nx.DiGraph()  #DiGraph showing spontaneous transitions 
                  #(no interactions between indivdiuals required)
H.add_node('SS')  #we actually don't need to include the 'SS' node in H.
H.add_edge('SI', 'SR', rate = 1) #An individual who is susceptible to disease
                                 #1 and infected with disease 2 will recover
                                 #from disease 2 with rate 1.
H.add_edge('IS', 'RS', rate = 1)
H.add_edge('II', 'IR', rate = 0.5)
H.add_edge('II', 'RI', rate = 0.5)
H.add_edge('IR', 'RR', rate = 0.5)
H.add_edge('RI', 'RR', rate = 0.5)

#In the below the edge (('SI', 'SS'), ('SI', 'SI')) means an
#'SI' individual connected to an 'SS' individual can lead to a transition in 
#which the 'SS' individual becomes 'SI'.  The rate of this transition is 0.18.
#
#Note that \texttt{IR} and \texttt{RI} individuals are more infectious than other 
#individuals.
#
J = nx.DiGraph()    #DiGraph showing induced transitions (require interaction).
J.add_edge(('SI', 'SS'), ('SI', 'SI'), rate = 0.18)
J.add_edge(('SI', 'IS'), ('SI', 'II'), rate = 0.18)
J.add_edge(('SI', 'RS'), ('SI', 'RI'), rate = 0.18)
J.add_edge(('II', 'SS'), ('II', 'SI'), rate = 0.18)
J.add_edge(('II', 'IS'), ('II', 'II'), rate = 0.18)
J.add_edge(('II', 'RS'), ('II', 'RI'), rate = 0.18)
J.add_edge(('RI', 'SS'), ('RI', 'SI'), rate = 1)
J.add_edge(('RI', 'IS'), ('RI', 'II'), rate = 1)
J.add_edge(('RI', 'RS'), ('RI', 'RI'), rate = 1)
J.add_edge(('IS', 'SS'), ('IS', 'IS'), rate = 0.18)
J.add_edge(('IS', 'SI'), ('IS', 'II'), rate = 0.18)
J.add_edge(('IS', 'SR'), ('IS', 'IR'), rate = 0.18)
J.add_edge(('II', 'SS'), ('II', 'IS'), rate = 0.18)
J.add_edge(('II', 'SI'), ('II', 'II'), rate = 0.18)
J.add_edge(('II', 'SR'), ('II', 'IR'), rate = 0.18)
J.add_edge(('IR', 'SS'), ('IR', 'IS'), rate = 1)
J.add_edge(('IR', 'SI'), ('IR', 'II'), rate = 1)
J.add_edge(('IR', 'SR'), ('IR', 'IR'), rate = 1)


return_statuses = ('SS', 'SI', 'SR', 'IS', 'II', 'IR', 'RS', 'RI', 'RR')

initial_size = 250 
IC = defaultdict(lambda: 'SS')
for individual in range(initial_size):   #start with some people having both
    IC[individual] = 'II'
for individual in range(initial_size, 5*initial_size): #and more with only 
                                                       #the 2nd disease
    IC[individual] = 'SI'

print('doing Gillespie simulation')
t, SS, SI, SR, IS, II, IR, RS, RI, RR = EoN.Gillespie_simple_contagion(G, H, 
                                                    J, IC, return_statuses,
                                                    tmax = float('Inf'))

plt.semilogy(t, IS+II+IR, '-.', label = 'Infected with disease 1')
plt.semilogy(t, SI+II+RI, '-.', label = 'Infected with disease 2')

plt.xlabel('$t$')
plt.ylabel('Number infected')
plt.legend()
plt.show()
\end{pyverbatim}
This produces a (stochastic) figure like

\includegraphics[width=0.8\columnwidth]{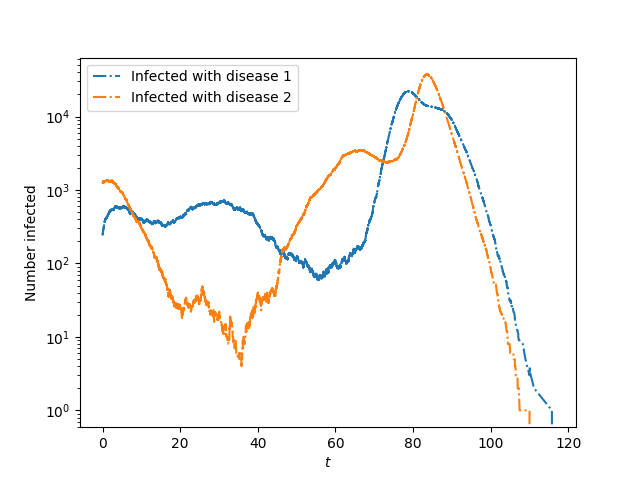}

\subsection*{Complex contagions}

Complex contagions are implemented through \texttt{Gillespie\_complex\_contagion} which 
allows a user to specify the rules governing a relatively arbitrary complex 
contagion.  The one criteria we note is that there is no memory - an individual 
will change from one status to another based on the current statuses of its 
neighbors, and not based on previous interactions with some neighbors who may 
have since changed status.

In the Gillespie implementation, we need a user-defined function which 
calculates the rate at which \texttt{u} will change status (given knowledge about 
the current state of the system) and another user-defined function which
chooses the new status of \texttt{u} given that it is changing status.  We finally 
need a user-defined function that will determine which other nodes have their 
rate change due to \texttt{u}'s transition.  By knowing the rates of all nodes
the Gillespie algorithm can choose the time of the next transition and which
node transitions.  Then it finds the new state, and finally it calculates the
new rates for all nodes affected by the change.

Once these functions are defined, the Gillespie algorithm is able to perform
the complex contagion simulation.

\subsubsection*{Example}

Previous work~\cite{miller:contagion} considered a dynamic version of the Watts 
Threshold Model~\cite{watts:WTM} spreading through clustered and unclustered 
networks.  The Watts Threshold Model is like an SI model, except that nodes 
have a threshold and must have more than some threshold number of infected 
partners before becoming infected.  The dynamic model in~\cite{miller:contagion} 
assumed that nodes transmit independently of one another, and a recipient 
accumulates transmissions until reaching a threshold and then switches status.  
An individual \texttt{v} can only transmit once to a partner \texttt{u}.  Because 
individuals cannot transmit to the same partner more than once it becomes nontrivial
to implement this in a way consistent with the 
memoryless criterion.  

Here we use another dynamic model that yields the same final state.  Once a 
node has reached its threshold number of infected partners, it transitions at 
rate 1 to the infected state.  The dynamics are different, but it can be proven that the
final states in both models are identical and follow deterministically from the
initial condition.  The following will produce the equivalent of Fig. 2a of~\cite{miller:contagion} for our new dynamic model. In that Figure, the threshold
was \texttt{2}.

\begin{pyverbatim}
import networkx as nx
import EoN
import numpy as np
import matplotlib.pyplot as plt
from collections import defaultdict

def transition_rate(G, node, status, parameters):
    '''This function needs to return the rate at which \texttt{node} changes status.
    For the model we are assuming, it should return 1 if \texttt{node} has at least
    2 infected partners and 0 otherwise.  The information about the threshold
    is provided in the tuple \texttt{parameters}.
    '''
    
    r = parameters[0] #the threshold
    
    #if susceptible and at least \texttt{r} infected partners, then rate is 1
    
    if status[node] == 'S' and len([nbr for nbr in G.neighbors(node) if 
                                    status[nbr] == 'I'])>=r:
        return 1
    else:
        return 0

def transition_choice(G, node, status, parameters):
    '''this function needs to return the new status of node.  We assume going
    in that we have already calculated it is changing status.
    
    this function could be more elaborate if there were different
    possible transitions that could happen.  However, for this model,
    the 'I' nodes aren't changing status, and the 'S' ones are changing to 
    'I'.  So if we're in this function, the node must be 'S' and becoming 'I'
    '''
    
    return 'I'
    
def get_influence_set(G, node, status, parameters):
    '''this function needs to return a set containing all nodes whose rates 
    might change because \texttt{node} has just changed status.  That is, which 
    nodes might \texttt{node} influence?
    
    For our models the only nodes a node might affect are the susceptible 
    neighbors.
    '''
    
    return {nbr for nbr in G.neighbors(node) if status[nbr] == 'S'}
    
parameters = (2,)   #this is the threshold.  Note the comma.  It is needed
                    #for python to realize this is a 1-tuple, not just a 
                    #number.   \texttt{parameters} is sent as a tuple so we need 
                    #the comma.
                    
N = 600000
deg_dist = [2, 4, 6]*int(N/3)
print('generating graph G with {} nodes'.format(N))
G = nx.configuration_model(deg_dist)
        
    
for rho in np.linspace(3./80, 7./80, 8):   #8 values from 3/80 to 7/80.
    print(rho)
    IC = defaultdict(lambda: 'S')
    for node in G.nodes():
        if np.random.random()<rho:  #there are faster ways to do this random 
                                    #selection
            IC[node] = 'I'
    
    print('doing Gillespie simulation')
    t, S, I = EoN.Gillespie_complex_contagion(G, transition_rate, 
                            transition_choice, get_influence_set, IC, 
                            return_statuses = ('S', 'I'), 
                            parameters = parameters)
                            
    print('done with simulation, now plotting')
    plt.plot(t, I)
    
plt.xlabel('$t$')
plt.ylabel('Number infected')
plt.show()
\end{pyverbatim}

This produces the (stochastic) figure\\
\includegraphics[width=0.8\columnwidth]{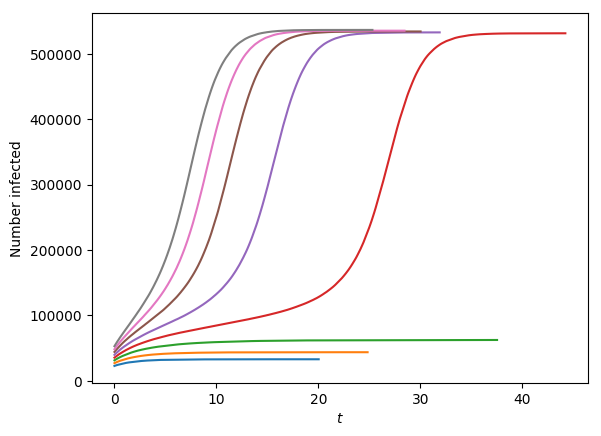}\\
which shows that if the initial proportion ``infected''
is small enough the final size is comparable to the initial size.  However
once the initial proportion exceeds a threshold, a global cascade occurs and 
infects almost every individual.

If we instead define \texttt{G} by 

\begin{pyverbatim}
deg_dist = [(0,1), (0,2), (0,3)]*int(N/3)
print('generating graph G with {} nodes'.format(N))
G = nx.random_clustered_graph(deg_dist)
\end{pyverbatim}

\texttt{G} will be a random clustered network~\cite{miller:random_clustered,newman:cluster_alg}, 
with the same degree distribution as before.  If we use a different range of values 
of \texttt{rho}, such as

\begin{pyverbatim}
for rho in np.linspace(1./80, 5./80, 8):
\end{pyverbatim}
this will produce a figure similar to Fig. 8a of~\cite{miller:contagion}.  Note that
the threshold initial size required to trigger a cascade is smaller in this
clustered network.

\includegraphics[width=0.8\columnwidth]{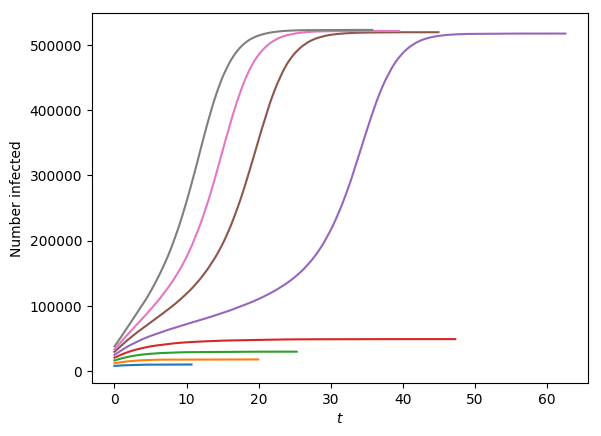}

\subsection*{Visualization \& Analysis}

By default the simulations return numpy arrays providing the number of individuals
with each state at each time.  However if we set a flag \texttt{return\_full\_data=True},
then the simulations return a \texttt{Simulation\_Investigation} object.  With the
\texttt{Simulation\_Investigation} object, there are methods which allow us to 
reconstruct all details of the simulation.  We can know the exact status of 
each individual at each time, as well as who infected whom.  

There are also 
methods  provided to produce output from the \texttt{Simulation\_Investigation} 
object.  These allow us to produce a snapshot of the network at a given time.
By default the visualization also includes the time series (e.g., S, I, and R) 
plotted beside the network snapshot.  These time series plots 
can be removed, or replaced by other time series, for example we could plot
multiple time series in the same axis, or time series generated by one of the
differential equations models.  With appropriate additional packages needed for
matplotlib's animation tools, the software can produce animations as well.

For SIR outbreaks, the \texttt{Simulation\_Investigation} object includes a 
transmission tree.  For SIS and simple contagions, it includes a directed 
multigraph showing the transmissions that occurred (this may not be a tree).
However for complex contagions, we cannot determine who
is responsible for inducing a transition, so the implementation does not provide
a transmission tree.  The transmission tree is useful for constructing synthetic
phylogenies as in~\cite{moshiri2018favites}.

\subsubsection*{Example - a snapshot of dynamics and a transmission tree for SIR disease.}

Using the tools provided, it is possible to produce a snapshot of the spreading
process at a given time as well as an animation of the spread.  We consider
SIR disease spreading in the Karate Club network~\cite{zachary1977information}.

\begin{pyverbatim}
import networkx as nx
import EoN
import matplotlib.pyplot as plt

G = nx.karate_club_graph()

nx_kwargs = {"with_labels":True} #optional arguments to be passed on to the 
                                 #networkx plotting command.
print('doing Gillespie simulation')
sim = EoN.Gillespie_SIR(G, 1, 1, return_full_data=True)
print('done with simulation, now plotting')

sim.display(time = 1, **nx_kwargs) #plot at time 1.
plt.show()
\end{pyverbatim}
This produces a (stochastic) snapshot at time 1: 

\includegraphics[width=0.8\columnwidth]{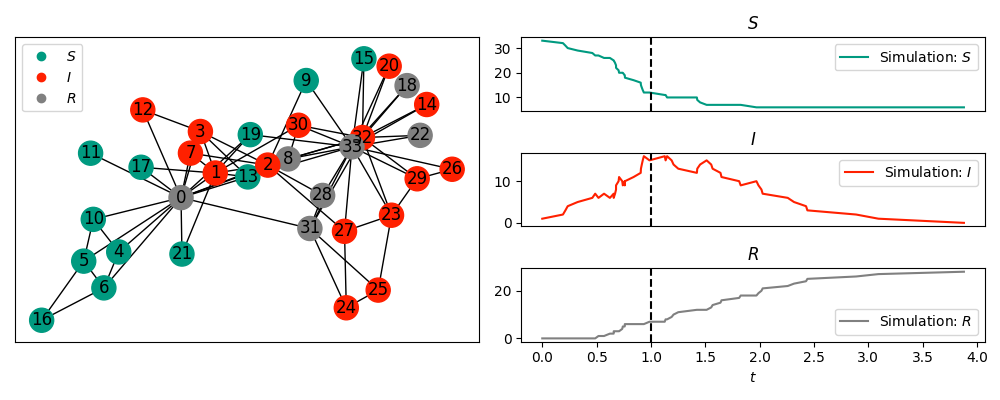}.

We can access the transmission tree.
\begin{pyverbatim}
T = sim.transmission_tree() #A networkx DiGraph with the transmission tree
Tpos = EoN.hierarchy_pos(T) #pos for a networkx plot

fig = plt.figure(figsize = (8,5))
ax = fig.add_subplot(111)
nx.draw(T, Tpos, ax=ax, node_size = 200, with_labels=True)
plt.show()
\end{pyverbatim}
This plots the transmission tree:

\includegraphics[width=0.8\columnwidth]{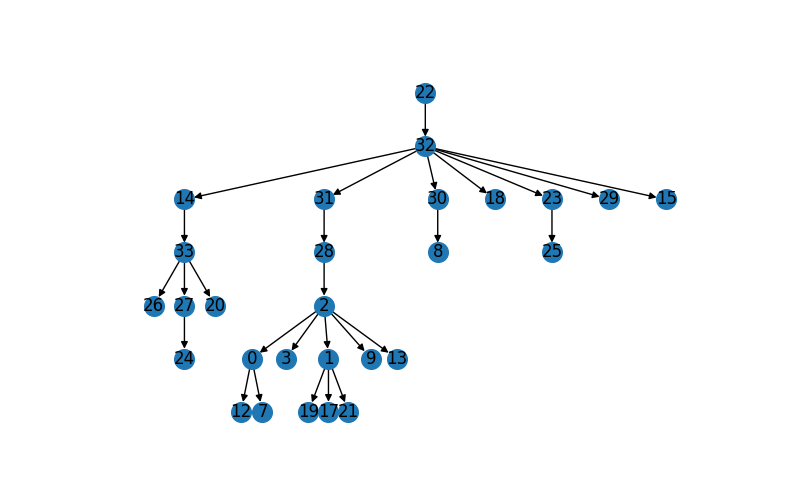}.

The command \texttt{hierarchy\_pos} is based on~\cite{stackoverflow:29586520}.

\subsubsection*{Example - Visualizing dynamics of SIR disease with vaccination.}

We finally consider an SIRV disease, that is an SIR disease with vaccination. 
As the disease spreads susceptible individuals get vaccinated randomly, without 
regard for the status of their neighbors.  

To implement this with EoN, we must use \texttt{Gillespie\_simple\_contagion}.  

We provide an animation showing the spread.  To make it easier to visualize, 
we use a lattice network.

\begin{pyverbatim}
import networkx as nx
import EoN
import matplotlib.pyplot as plt
from collections import defaultdict

print('generating graph G')
G = nx.grid_2d_graph(100,100) #each node is (u,v) where 0<=u,v<=99
#we'll initially infect those near the middle 
initial_infections = [(u,v) for (u,v) in G if 45<u<55 and 45<v<55]

H = nx.DiGraph()  #the spontaneous transitions
H.add_edge('Sus', 'Vac', rate = 0.01)
H.add_edge('Inf', 'Rec', rate = 1.0)

J = nx.DiGraph()  #the induced transitions
J.add_edge(('Inf', 'Sus'), ('Inf', 'Inf'), rate = 2.0)

IC = defaultdict(lambda:'Sus') #a "dict", but by default the value is \texttt{'Sus'}.
for node in initial_infections:
    IC[node] = 'Inf'
    
return_statuses = ['Sus', 'Inf', 'Rec', 'Vac']

color_dict = {'Sus': '#009a80','Inf':'#ff2000', 'Rec':'gray','Vac': '#5AB3E6'}
pos = {node:node for node in G}
tex = False
sim_kwargs = {'color_dict':color_dict, 'pos':pos, 'tex':tex}

print('doing Gillespie simulation')
sim = EoN.Gillespie_simple_contagion(G, H, J, IC, return_statuses, tmax=30, 
                            return_full_data=True, sim_kwargs=sim_kwargs)

times, D = sim.summary() 
#
#times is a numpy array of times.  D is a dict, whose keys are the entries in
#return_statuses.  The values are numpy arrays giving the number in that 
#status at the corresponding time.
                      
newD = {'Sus+Vac':D['Sus']+D['Vac'], 'Inf+Rec' : D['Inf'] + D['Rec']}
#
#newD is a new dict giving number not yet infected or the number ever infected
#Let's add this timeseries to the simulation.
#
new_timeseries = (times, newD) 
sim.add_timeseries(new_timeseries, label = 'Simulation', 
                    color_dict={'Sus+Vac':'#E69A00', 'Inf+Rec':'#CD9AB3'})

sim.display(time=6, node_size = 4, ts_plots=[['Inf'], ['Sus+Vac', 'Inf+Rec']])
plt.show()
\end{pyverbatim}
This plots the simulation at time 6.

\includegraphics[width=0.8\columnwidth]{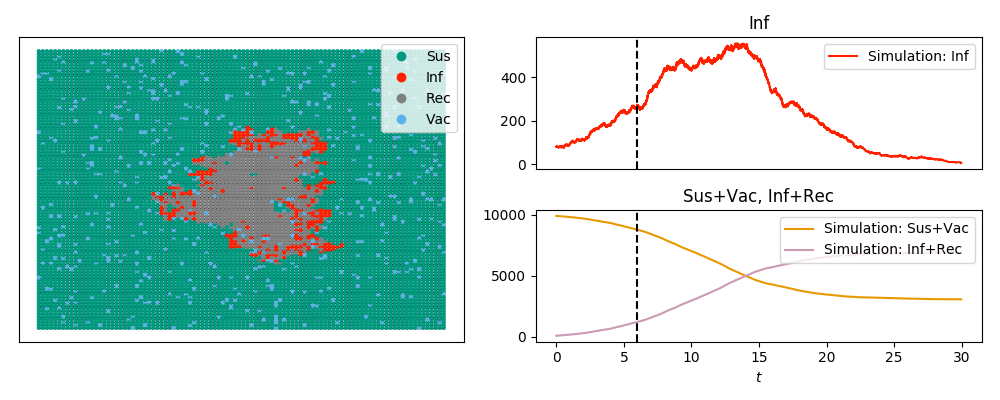}

We can also animate it

\begin{pyverbatim}
ani=sim.animate(ts_plots=[['Inf'], ['Sus+Vac', 'Inf+Rec']], node_size = 4)  
ani.save('SIRV_animate.mp4', fps=5, extra_args=['-vcodec', 'libx264'])
\end{pyverbatim}

This will create an mp4 file animating previous display over all calculated 
times.  Depending on the computer installation, \texttt{extra\_args} will need to be
modified.

\section*{Discussion}

EoN provides a number of tools for studying infectious processes spreading in
contact networks.  The examples given here are intended to demonstrate the
range of EoN, but they represent only a fraction of the possibilities.

Full documentation is available at 
\begin{center}
\href{https://epidemicsonnetworks.readthedocs.io/en/latest/}{https://epidemicsonnetworks.readthedocs.io/en/latest/}.
\end{center}

\section*{Dependencies}

scipy
numpy
networkx
matplotlib

\section*{Related Packages}

There are several alternative software packages that allow for simulation of 
epidemics on networks.  Here we briefly review some of these.

\subsection*{epydemic}

Epydemic is a python package that can simulate SIS and SIR epidemics in 
networks.  It is also built on networkx.  It can handle both discrete-time 
simulations or continuous-time Markovian simulations for which it uses a 
Gillespie-style algorithm.  It can handle more processes than just SIS or SIR 
disease.  In fact it can handle any model which can be simulated using the 
\texttt{EoN.simple\_contagion}.

The documentation is available at 
\href{https://pyepydemic.readthedocs.io/en/latest/}{https://pyepydemic.readthedocs.io/en/latest/}.

\subsection*{Graph-tool}

Graph-tool~\cite{peixoto_graph-tool_2014} is a python package that serves as an 
alternative to networkx.  Many of its underlying processes are written in C++, 
so it is often much faster than networkx.

Graph-tool has a number of built-in dynamic models, including the SIS, SIR, 
and SIRS models.  The disease models are currently available only in 
discrete-time versions.

The documentation for these disease models is available at 
\begin{center}
\href{https://graph-tool.skewed.de/static/doc/dynamics.html}{https://graph-tool.skewed.de/static/doc/dynamics.html}.
\end{center}

\subsection*{EpiModel}

EpiModel~\cite{jenness:EpiModel} is an R package that can handle SI, SIS, and SIR 
disease spread.  It is possible to extend EpiModel to other models.  EpiModel
is built around the StatNet package.  More details about EpiModel are available
at 
\begin{center}
\href{https://www.epimodel.org/}{https://www.epimodel.org/}.
\end{center}

\section*{Acknowledgments}

The development of EoN has been supported by Global Good and by La Trobe 
University.  The inclusion of python code in this paper was facilitated by the package pythontex~\cite{poore2015pythontex}.  

\bibliographystyle{plain}
\bibliography{paper}

\end{document}